\documentclass[prl,twocolumn,showpacs]{revtex4}
\usepackage{amsmath,amssymb,mathrsfs}
\usepackage{psfrag}
\usepackage{graphicx}
\usepackage{graphics}
\usepackage{epsfig}
\usepackage{bm}
\usepackage{color}
\usepackage{verbatim,color,ulem}

\begin{document}

\title{Effects of long range hopping in the Bose-Hubbard model}
\author{M. Ferraretto$^1$ and L. Salasnich$^{1,2}$} 
\affiliation{$^1$Dipartimento di Fisica e Astronomia ``Galileo Galilei'', 
Universit\`a di Padova, Via Marzolo 8, 35131 Padova, Italy
\\
$^2$Istituto Nazionale di Ottica (INO) del Consiglio Nazionale 
delle Ricerche (CNR), Via Nello Carrara 1, 50019 Sesto Fiorentino, Italy}

\date{\today}

\begin{abstract}
We investigate the effects of an extended Bose-Hubbard model with 
a long range hopping term on the Mott insulator-superfluid quantum 
phase transition. We consider the effects of a power law 
decaying hopping term and show that the Mott phase is shrinked in 
the parameters' space. We provide an exact solution for one dimensional 
lattices and then two approximations for higher dimensions, each one valid 
in a specific range of the power law exponent: a continuum approximation and 
a discrete one. Finally, we extend these results to a more realistic 
situation, where the long range hopping term is made by a power law 
factor and a screening exponential term and study the main effects 
on the Mott lobes. 
\end{abstract}

\pacs{03.75.Ss 03.70.+k 05.70.Fh 03.65.Yz} 

\maketitle

\section{I. Introduction} 

After the experimental realization of Bose-Einstein condensation 
in 1995 \cite{Cornell,Hulet,Ketterle}, the research in ultracold 
atomic physics has known great developement. 
In particular, theoretical and experimental efforts have been done to study 
quantum phases of ultracold atomic bosons in optical lattices \cite{Zoller}. 
In this context, the pioneer experiment was performed by 
Greiner {\it et al.} in 2002 \cite{Bloch}. The authors confined a 
Bose-Einstein condensate with repulsive interactions at very 
low temperature in a three dimensional optical lattice and studied 
the interference pattern produced by the system as function of the lattice 
depth. A transition between a superfluid phase (each atom is spread out 
over the entire lattice and the interference pattern is peaked at reciprocal 
lattice wavevectors) and a Mott insulator phase (with a precise number 
of atoms in every well and a gaussian interference pattern peaked at 
zero-wavevector) is observed at some critical depth \cite{Zwerger}.

The many-body model which describes this phase transition is the 
Bose-Hubbard Hamiltonian 
\begin{equation}
\label{eq:BoseHubbardHamiltonian}
\hat{H} = - J \sum_{\left\langle ij \right\rangle} \hat{b}_i^{\dagger}\hat{b}_j 
+ \frac{U}{2}\sum_i \hat{n}_i \left( \hat{n}_i - 1 \right)
\end{equation}
where $J$ is the hopping energy, $U$ is the interaction energy 
(both are assumed positive), $\hat{b}_i$ and $\hat{b}^{\dagger}_i$ 
are on-site bosonic annihilation and creation operators, 
$\hat{n}_i=\hat{b}^{\dagger}_i \hat{b}_i$ and $\left\langle ij \right\rangle$ 
means that $i$ and $j$ are nearest neighbors lattice sites.
The effective parameters $J$ and $U$ can be obtained by microscopic 
quantities (mass and scattering length of atoms) and lattice parameters 
(spacing and depth) introducing Wannier functions, as shown in 
Refs. \cite{Zoller, Bloch, Zwerger}.

In recent years, many authors have extended the standard model 
to investigate new quantum phases, new transitions, and other properties 
with theoretical means and Quantum Monte Carlo simulations. In particular, 
some remarkable studies have considered exotic geometries, 
such as Bethe lattices, complex networks and more \cite{Extension, 
Extension1, Extension2}. Some others have taken into account 
more complicated interaction terms, such as nearest and next nearest 
neighbors interactions, both for spinless and spin-1 bosons 
\cite{Extension3, Extension4, Extension5}. Even more recently, 
quantum phase transitions in disordered systems have been 
investigated \cite{Extension6, Extension7}.

In the present work we will consider lattices with a simple geometry 
(hypercubic lattice in $d$ dimensions) and ordered systems with only 
on-site interactions between atoms. Instead, we will take into account 
the possibility for an atom to tunnel from a site $i$ to every 
other site $j$ and study the effect of this generalization 
on the phase diagram.

\section{II. Landau effective action}

Using a path integral approach, as discussed in Refs. 
\cite{Dupuis, Pelster}, an effective action for the theory can be obtained 
after a Hubbard-Stratonovic decoupling of the hopping term. 
The relevant aspect for the present work is that this procedure 
can be set up in a more general frame, namely taking a more general 
hopping, such that our generalized Bose-Hubbard Hamiltonian reads 
\begin{equation}
\hat{H} = -\sum_{ij}J_{ij}\hat{b}^{\dagger}_i\hat{b}_j 
+ \frac{U}{2}\sum_i \hat{n}_i \left( \hat{n}_i - 1 \right)
\end{equation}
where the sum over $i$ and $j$ is not restricted to nearest neighbors. 

Close to the superfluid-Mott phase transition, the order parameter 
$\psi({\vec r},t) = \langle {\hat b}_i(\tau) \rangle$ of the system 
corresponds to the expectation value of the bosonic annihilation 
operator ${\hat b}_i$ at the imaginary time $\tau$ and at the site 
$i$ associated to the spatial position ${\vec r}$ \cite{Dupuis, Pelster}. 
After Fourier transforming this field and ignoring the effect of 
its fluctuations ($\psi_{\vec{q}}=0$ for every momentum $\vec{q} \neq 0$), 
one obtains the effective mean-field Landau action \cite{Dupuis, Pelster}.
\begin{equation}
\label{eq:LandauAction} 
S[\psi_0^{\dagger},\psi_0] = S_0 + c_2 \left| \psi_0 \right|^2 + 
c_4 \left| \psi_0 \right|^4
\end{equation}
where $c_4>0$ and $\psi_0$ is the order parameter. 
This action has only one global minimum $\psi_0=0$ when $c_2>0$, 
while develops infinite equivalent minima given by $\left| \psi_0 \right| 
= \sqrt{-c_2/2c_4}$ when $c_2<0$. Since in the second case the $U(1)$ symmetry 
of the action in Eq. (\ref{eq:LandauAction}) is spontaneously broken by 
the ground state, we interprete this as the superfluid phase, 
$\left| \psi_0 \right|^2$ being the superfluid density, and the other case 
as the Mott insulating phase. The critical behaviour occurs when $c_2=0$, 
the coefficient $c_2$ being
\begin{equation}
c_2 = \beta N^d \left(J^{-1}_0 - G_{loc}(0) \right)
\end{equation}
where $N^d$ is the total number of lattice sites ($N$ in each direction), 
$\beta$ is the inverse temperature, $J_0$ is the zero-momentum Fourier 
coefficient of the interaction matrix
\begin{equation}\label{eq:Jq}
J_{\vec{q}} = \frac{1}{N} \sum_{ij} J_{ij} 
e^{i\vec{q}\cdot \left(\vec{r}_i-\vec{r}_j \right)}
\end{equation}
and $G_{loc}(0)$ is the local Green function at zero frequency
\begin{equation}\label{eq:Gloc}
G_{loc}(0) = \frac{\bar{n}}{\mu - U(\bar{n}-1)} - 
\frac{\bar{n}+1}{\mu - U\bar{n}}
\end{equation}
In Eq. (\ref{eq:Gloc}) $\mu$ is the chemical potential and 
$\bar{n}=\lceil \mu / U \rceil$ is the first integer greater than 
$\mu / U$ and represents the number of atoms in every site in the Mott 
phase (the lattice is not empty only when $\mu>0$). For our purposes, 
the most important property of the local Green function is that 
$G_{loc}(0)>0$ for every $\mu$.

For nearest neighbors hopping with $J_{ij}=J$ 
and hypercubic lattices in $d$ dimensions, 
writing $\vec{q} = \sum_{k=1}^d 2\pi q_k \hat{e}_k/L$ (where $q_k \in 
\mathbb{Z}$ and $\hat{e}_k$ is the $k$-th element of the canonic base) 
one can perform the sums in Eq. (\ref{eq:Jq}) to find
\begin{equation}
J_{\vec{q}} = 2J\sum_{k=1}^d \cos{ \left( \frac{2\pi a}{L}q_k \right)}
\end{equation}
Hence the zero momentum term is $J_0 = 2dJ$ and the coefficient $c_2$ is
\begin{equation}\label{eq:c2NN}
c_2 = \frac{1}{2dJ} - G_{loc}(0)
\end{equation}

\section{III. Bose-Hubbard model with long range hopping} 

The main purpose of this work is to extend the approach discussed in the 
previous section to a generalized Bose-Hubbard model 
in $d$ dimentional hypercubic 
lattices with power law 
decaying hopping energy of the form
\begin{equation}\label{eq:PowerLawJ}
J_{ij} = \frac{J a^s}{\left|\vec{r}_i - \vec{r}_j\right|^s}
\end{equation}
where $a$ is the lattice spacing and hence $J$ is still the nearest neighbors 
hopping energy. The exponent $s$ defines the hopping range: the larger $s$, 
the smaller the probability of long range hopping. In the limit 
$s \rightarrow \infty$ the hopping is $J$ when $i$ and $j$ are nearest 
neighbors and $0$ otherwise, which is the case studied in the previous section.

The zero momentum coefficient $J_0$ can be found setting $\vec{q}=0$ in 
Eq. (\ref{eq:Jq}) and solving $J_0 = (1/N)\sum_i \sum_{j\neq i} J_{ij}$. 
Taking periodic boundary conditions, every site of the lattice can be 
regarded as a bulk site, no surface effects have to be taken into account 
and then there is a perfect discrete translational invariance: this means 
that the sum $\sum_{j \neq i} J_{ij}$ does not depend on $i$. As a consequence, 
the sum over $i$ can be performed and gives a factor $N$, so that 
$J_0 = \sum_{j\neq 0} J_{0j}$.

Following the same procedure performed in Refs. \cite{Ising, Ising2} 
for the Ising model, the remaining sum can be evaluated using the continuum 
approximation with the prescription $a^d \sum_i \rightarrow \int d^d r$. 
This is consistent for not too large values of $s$, since in this way the 
integrand function doesn't change abruptly from one site to his neighbors.
\begin{equation}
J_0 = \frac{J}{a^{d-s}} \int d^d r \frac{1}{r^s}
\end{equation}
The above integration is performed using polar coordinates in $d$ dimensions 
($d^dr = r^{d-1}d\Omega_d dr$), since the angular part can be integrated and 
gives the $d$ dimensional solid angle
\begin{equation}
\Omega_d = \frac{2\pi^{d/2}}{\Gamma(d/2)}
\end{equation}
where $\Gamma(x)$ is the Euler Gamma function.

To compute the radial part we introduce upper and lower physical cutoffs. 
The most reasonable lower cutoff is the lattice spacing $a$, while the upper 
one is the lattice size length $Na$ \cite{Ising2}.
\begin{equation}\label{eq:J0}
\begin{array}{ccc}
J_0 & = & \frac{J\Omega_d}{a^{d-s}} \int_a^{Na} r^{d-s-1} dr \\
 &   & \\
 & = & 
\begin{cases}
\frac{J\Omega_d}{d-s} \left( N^{d-s}-1 \right) & \mbox{ if }  s \neq d \\
J\Omega_d \log{N} &  \mbox{ if }  s = d
 \end{cases}
 \end{array}
\end{equation}

Taking the thermodynamic limit $N \rightarrow \infty$ we can compute $J_0^{-1}$ 
and finally write the coefficient $c_2$:
\begin{equation}\label{eq:c2}
c_2 = \begin{cases}
-G_{loc}(0) & \mbox{ if } s \leq d \\
\frac{s-d}{J\Omega_d}-G_{loc}(0) & \mbox{ if } s > d
\end{cases}
\end{equation}

By comparing Eqs. (\ref{eq:c2NN}) and (\ref{eq:c2}) we can formally find 
the exponent $s_0$ for which the two equations are equal
\begin{equation}\label{eq:s0}
s_0 = d + \frac{\pi^{d/2}}{d \Gamma(d/2)}
\end{equation}

Since the nearest neighbor hopping should be recovered in the limit 
$s\rightarrow \infty$ (and not $s \rightarrow s_0$), we can use the value 
$s_0$ as an upper limit for our approximation and expect it to give good 
results only for $s \ll s_0$.

\begin{figure}
\includegraphics[scale=0.6]{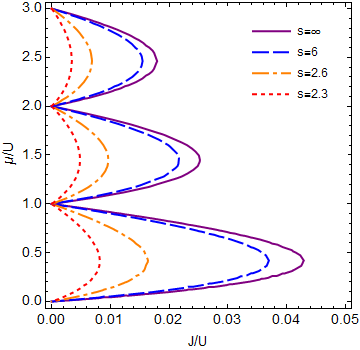}
\caption{Boundary between the two phases 
(Mott phase inside the lobes and superfluid outside) in a two dimensional 
lattice, for some values of $s$. The lobes corresponding to $s=2.3$ 
and $s=2.6$ have been computed with the continuum approximation; 
the lobe at $s=6$ has been computed with the discrete approximation 
of Eq. (\ref{eq:Exact2d}). Finally, the lobe at $s=\infty$ corresponds 
to nearest neighbors hopping. As $s$ moves from $\infty$ to $d$ the lobes 
are shrinked and eventually disappear for $s=d$.}
\label{fig:MottLobes}
\end{figure}

The critical line at zero temperature in the parameters space $J/U$, $\mu /U$ 
is given by the condition $c_2=0$.
When the range exponent $s$ is greater than the critical value $d$ 
(i.e. when the hopping is short-ranged), the phase diagram has the same 
shape as the one obtained with nearest-neighbors hopping, even if the phase 
boundary is shrinked depending on $s$. The lobes at $s=2.3$, $s=2.6$ 
in Fig. \ref{fig:MottLobes} have been computed within this approximation 
in $d=2$.

When $s$ is below his critical value $d$ (i.e. when the hopping is 
long ranged), the thermodynamic limit is not well defined since the system 
would have infinite energy. However, if $N$ is finite but large, 
since $G_{loc}(0)$ is positive, $c_2 \approx -G_{loc}(0)$ is negative and 
the Mott-phase is almost unavailable for the system.

\section{IV. Exact solution in one dimensional lattice}

If the dimensionality of the system is $d=1$, the continuum approximation 
is not necessary since there is a simple exact solution. 
Taking $i=0$ and $\vec{r}_j = aj$, $j \in \mathcal{Z}$, the sum we have 
to perform is $\sum_{j\neq 0} 1/|j|^s$. This sum diverges for $s \leq 1$ 
and converges to $2\zeta(s)$ (where $\zeta(s)$ is the Riemann zeta function) 
otherwise; so the coefficient $c_2$ is
\begin{equation}
\label{eq:Exact1d}
c_2 = \begin{cases}
-G_{loc}(0) & \mbox{ if } s \leq 1 \\
\frac{1}{2J\zeta(s)} - G_{loc}(0) & \mbox{ if } s>1
\end{cases}
\end{equation}

\begin{figure}
\includegraphics[scale=0.5]{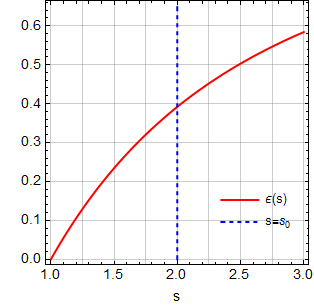}
\includegraphics[scale=0.5]{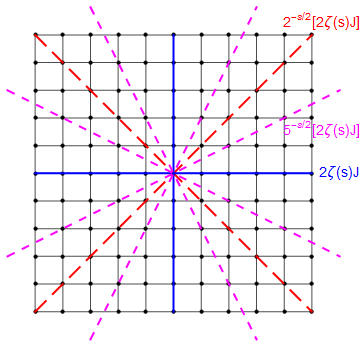}
\caption{Upper panel: the curve represents the error $\varepsilon(s)$ 
of the continuous approximation in $d=1$; the dashed line shows $s_0=2$, 
which is the limit for the approximation. When $s<s_0$, the error is below 
the 40\%. Lower panel: visual picture of Eq. \ref{eq:Exact2d} 
in a square lattice in $d=2$. Summing the contributions given by 
all sites in a line, we get a contribute proportional to $2\zeta(s)J$.}
\label{fig:1d-Precision}
\end{figure}

In the limit $s \rightarrow 1^+$, since $\zeta(s) \approx (s-1)^{-1}$, 
this result is in good agreement with the continuum approximation 
of Eq. (\ref{eq:c2}).  In the opposite limit $s \gg s_0$, 
where $s_0=2$ the approximation is expected to be unprecise. 
In Fig. \ref{fig:1d-Precision} we present the continuum approximation 
error $\varepsilon(s)=1-J^{approx}_0/J_0^{true}=1-[(s-1)\zeta(s)]^{-1}$ 
as function of $s$ to give a visual picture of what we have stated above.

It is remarkable that, as stated in Refs. \cite{1d} and \cite{Review}, 
the mean field theory fails at describing the Mott-superfluid 
transition in $d=1$. The exact mean field solution provided in this 
section does not predict the correct shape of the Mott lobes, 
but is a useful test for the accuracy of our approximation.

The exact solution of the one dimensional model suggests generalization 
for higher dimensional hypercubic lattices. For example, in a $d=2$ square 
lattice, where $j$ is labeled by two integer indices $m,n$, we have to 
sum $(m^2+n^2)^{-s/2}$ over the whole lattice excluding the origin. 
The contribution of all the sites lying on a straight line passing through 
the origin is proportional to $2\zeta(s)J$. The proportionality constant 
is the inverse distance between two sites on that line to the power of $s$, 
namely $(p^2+q^2)^{-s/2}$, where $p$ and $q$ are the smallest coordinates 
of a point in that line. We can span one fourth of the lattice taking 
straight lines with slope only between $0$ and $1$ and notice that, 
by symmetry, every contribution is repeated $4$ times, except the ones 
due to slope $0$ and slope $1$ lines, which are only repeated twice. 
The proportionality factor is $1$ for the horizontal line and $2^{-s/2}$ 
for the bisector; while for all the other lines it is $(p^2+q^2)^{-s/2}$ 
with $0<q<p$ and $q$ coprime to $p$ (the sum over all values of $q$ 
respecting this conditions is indicated as $\sum'_q$). 
All this considerations lead to the exact equation
\begin{equation}\label{eq:Exact2d}
J_0 = 4\zeta(s)J \left[1 + 2^{-s/2} + 2\sum_{p=2}^{\infty} 
\sum_q' (p^2 + q^2)^{-s/2} \right] 
\end{equation}

The main idea is to use Eq. (\ref{eq:Exact2d}) as an approximation 
for high $s$ by cutting off the sum at some point. 
This provides a good approximation for $s \gg s_0$ (when the system is 
intrinsically discrete) and as we expect, in the limit 
$s \rightarrow \infty$, $J_0 \rightarrow 4J$, which is the result for 
nearest neighbors hopping. A picture of the scheme proposed above 
is provided in Fig. \ref{fig:1d-Precision}. In Fig. \ref{fig:MottLobes} we have 
computed the lobe at $s=6$ using this approximation.

\section{V. Long range hopping with a screening term}

In some contexts \cite{Illuminati} a more physical form of the 
interaction matrix is
\begin{equation}\label{eq:ExponentialJ}
J_{ij} = \frac{Ja^s}{|\vec{r}_i-\vec{r}_j|^s} e^{-|\vec{r}_i-\vec{r}_j|^2/4l^2}
\end{equation}
where $l$ is a characteristic length of the Wannier function 
which decreases when increasing the lattice depth. In this context 
a dimensionless control parameter 
\begin{equation}
\eta = {a\over 2l}  
\end{equation} 
can be defined: in the limit $\eta \rightarrow 0$ we recover 
the previous case.

If $J_{ij}$ doesn't change abruptly from one site to his 
neighbors ($s \ll s_0$, $\eta \ll 1$), we can apply the continuum 
approximation and use integrals instead of sums. The radial part 
of the integration can be performed with the substitution $t=r^2/4l^2$, 
introducing the same cutoffs and using the upper incomplete Euler Gamma 
function $\Gamma(z,x)=\int_x^{\infty} t^{z-1} e^{-t}dt$.
\begin{equation}\label{eq:Ferraretto}
J_0 = \frac{J\Omega_d}{2\eta^{d-s}} \left[\Gamma \left(\frac{d-s}{2} , 
\eta^2 \right) - \Gamma \left( \frac{d-s}{2}, N^2 \eta^2 \right) \right]
\end{equation}

For a fixed value of $N$, since for $\eta \rightarrow 0$ the function 
$\Gamma(z,\eta^2) - \Gamma(z,N^2\eta^2) \approx \eta^{2z}(N^{2z}-1)/z$ 
when $z\neq 0$, while $\Gamma(0,\eta^2) - \Gamma(0,N^2\eta^2) 
\approx 2 \log{N}$, in this limit we get Eq. (\ref{eq:J0}) as expected.

The thermodynamic limit is realized taking $\eta$ small but fixed 
and $N \rightarrow \infty$ in Eq. (\ref{eq:Ferraretto}). Considering that 
$\lim_{x \rightarrow \infty} \Gamma(z,x) = 0$, we conclude that in the 
thermodynamic limit $J_0$ is always finite and positive.
The corresponding coefficient $c_2$ is
\begin{equation}
c_2 = \frac{2\eta^{d-s}}{J \Omega_d \Gamma\left(\frac{d-s}{2}, 
\eta^2 \right) } - G_{loc}(0)
\end{equation}

This fact has remarkable consequences in the phase diagram of 
Fig. \ref{fig:MottLobes}, because the Mott phase is now available 
for the system for every $s$, even in the thermodynamic limit: 
the exponential attenuation factor significantly screens the hopping, 
lowering the effective range. 

\begin{figure}
\includegraphics[scale=0.6]{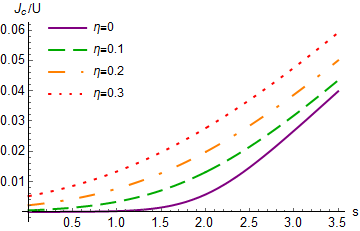}
\caption{Amplitude of the first Mott lobe ($J_c/U$) as a function 
of the exponent $s$ in a two dimensional lattice with $N=100$ 
at different values of $\eta$.}
\label{fig:CriticalJ}
\end{figure}

\section{VI. Conclusions}

In this work we have presented the main results of the standard 
Bose-Hubbard theory with hopping between nearest neighbors, and 
then we have extended the model considering tunneling from 
one site to every other, with hopping energy decreasing with 
the distance as a power law. We have studied the extended model 
using two approximations: a continuum approximation 
(substitution of sums with integrals) for $s \ll s_0$ and a 
discrete approximation for $s \gg s_0$, inspired by the exact 
solution to the one dimensional problem. 
In both cases we have assumed perfect discrete translational invariance 
along the directions of the basis vectors by taking periodic boundary 
conditions. We have checked the consistency of these approximations, 
and we have deduced the most important result in the thermodynamic limit: 
when the range exponent $s$ is lower than the lattice 
dimension $d$ the Mott phase does not exist; instead, 
above this critical value $s=d$ the phase transition occurs 
but the Mott phase is shrinked in the parameters space. 
Howeover, if the system is finite, the phase transition 
occurs even when $s<d$, but the Mott phase is extremely 
shrinked in parameters' space. Finally, we have considered 
a more general and physically motivated hopping energy, 
taking into account a screening exponential term with a control 
parameter $\eta$ and applying the continuum approximation, 
valid for small $s$ and small $\eta$. 
The screening term weakens the long range hopping and makes 
the phase transition possible for every $s$, even in the thermodynamic 
limit, which in this case is obtained taking $N \rightarrow \infty$ 
at fixed $\eta$. Moreover, increasing $\eta$ leads to an expansion 
of the Mott phase in the parameters' space.

\vspace{0.5cm}

The authors thank M. Faccioli and F. Toigo for useful discussions. 
L.S. acknowledges for partial support the FFABR grant of Italian 
Ministry of Education, University and Research.


\begin{thebibliography}{99}

\bibitem{Cornell} M.H. Anderson, J.R. Ensher, M.R. Matthews, 
C.E. Wieman, and E.A. Cornell, Science {\bf 269}, 198 (1995). 

\bibitem{Hulet} C. C. Bradley, C. A. Sackett, J. J. Tollett, and 
R. G. Hulet, Phys. Rev. Lett. {\bf 75}, 1687 (1995). 

\bibitem{Ketterle} K.B. Davis, M.-O. Mewes, M.R. Andrews, 
N.J. van Druten, D.S. Durfee, D.M. Kurn, and W. Ketterle, 
Phys. Rev. Lett. {\bf 75}, 3969 (1995). 

\bibitem{Zoller} D. Jaksch, C. Bruder, J. I. Cirac, C. W. Gardiner, 
and P. Zoller, Phys. Rev. Lett. \textbf{81}, 3108 (1998).

\bibitem{Bloch} M. Greiner, O. Mandel, T. Esslinger, T. W. Hänsch, 
and I. Bloch, Nature, \textbf{415}, 39–44 (2002).

\bibitem{Zwerger} I. Bloch, J. Dalibard, and W. Zwerger, Rev. Mod. Phys. 
\textbf{80}, 885 (2008).

\bibitem{Extension} G. Semerijan, M. Tarzia, and F. Zamponi, 
Phys. Rev. B \textbf{80}, 014524 (2009).

\bibitem{Extension1} A. Halu, L. Ferretti, A. Vezzani, and G. Bianconi, 
EPL \textbf{99}, 18001 (2012).

\bibitem{Extension2} R. Sachdeva, F. Metz, M. Singh, T. Mishra, 
and T. Busch, e-preprint arXiv:1808.05348.

\bibitem{Extension3} D. Rossini, and R. Fazio, 
New J. Phys. \textbf{14}, 065012 (2012).

\bibitem{Extension4} S. N. Nabi, and S. Basu, e-preprint 
arXiv:1705.00475.

\bibitem{Extension5} K. Biedron, M. Lacki, and J. Zakrzewski, 
Phys. Rev. B \textbf{97}, 245102 (2018).

\bibitem{Extension6} F. Lin, T. A. Maier, and V. W. Scarola, 
Sci. Rep. \textbf{7}, 12752 (2017).

\bibitem{Extension7} B. R. de Abreu, U. Ray, S. A. Vitiello, and 
D. M. Ceperley, Phys. Rev. A \textbf{98}, 023628 (2018).

\bibitem{Dupuis} K. Sengupta, and N. Dupuis, 
Phys. Rev. A \textbf{71}, 033629 (2005).

\bibitem{Pelster} B. Bradlyn, F. E. A. dos Santos, and 
A. Pelster, Phys. Rev. A \textbf{79}, 013615 (2009).

\bibitem{1d} S. Ejima, H. Fehske, F. Gebhard, K. zu Munster, M. Knap, 
E. Arrigoni, and W. von der Linden, Phys. Rev. A \textbf{85}, 053644 (2012).

\bibitem{Ising} E. Luijten, and H. W. J. Blote, 
Phys. Rev. B \textbf{56}, 8945 (1997).

\bibitem{Ising2} S. A. Cannas, and F. A. Tamarit, Phys. Rev. B 
\textbf{56}, R12661(R) (1996).

\bibitem{Illuminati} G. Mazzarella, S. M. Giampaolo, and 
F. Illuminati, Phys. Rev. A \textbf{73}, 013625 (2006).

\bibitem{Review2} M. Lewenstein, A. Sanpera, V. Ahufinger, B. Damski, 
A. Sen De, and U. Sen, Adv. in Phys. {\bf 56}, 243 (2007).

\bibitem{Review} K. V. Krutitsky, Phys. Rep. {\bf 607}, 1 (2016).

\end{thebibliography}
\end{document}